\documentstyle[preprint,aps]{revtex}
\def\lprox{\mathrel{\raise .3ex\hbox{$<$\kern-
.75em\lower1ex\hbox{$\sim$}}}}

\def\gprox{\mathrel{\raise .3ex\hbox{$>$\kern-
.75em\lower1ex\hbox{$\sim$}}}}

\begin{document}

\draft

\title{\bf Averaged Energy Conditions in 4D \\ Evaporating Black Hole 
Backgrounds}

\author{Hongwei Yu\footnote{e-mail: hwyu@cosmos2.phy.tufts.edu}}

\address{Institute of Cosmology, Department of Physics and Astronomy\\
Tufts University, Medford, MA 02155, USA}

%\date{}

\maketitle

\begin{abstract}

Using Visser's semi-analytical model for the stress-energy tensor corresponding to 
the conformally coupled massless scalar field in the Unruh vacuum, we examine,
 by explicitly evaluating the relevant integrals over half-complete geodesics,  
the averaged weak (AWEC) and averaged null (ANEC) energy conditions along with 
Ford-Roman quantum inequality-type restrictions on negative energy in the context 
of four dimensional evaporating black hole backgrounds. We find that in all cases
 where the averaged energy conditions fail, there exist quantum inequality bounds
 on the magnitude and duration of negative energy densities.

\end{abstract}

\section{ Introduction}

It is well-known nowadays that quantum field theory allows violations of
 all the known  pointwise energy conditions in classical general relativity, 
such as the weak energy 
condition and null energy condition \cite{HE}. 
Specific examples are the Casimir effect \cite{C,BM}, squeezed states 
of light 
\cite{WKHW} and 
black hole evaporation \cite{H75} which involves negative energy densities 
and fluxes in a crucial way. On the other hand, if the laws 
of quantum field theory place no restrictions on negative energy, then it 
might be 
possible to produce gross macroscopic effects such as: violation of the second 
law of thermodynamics \cite{F78,D82} or of cosmic censorship \cite{FR90,FR92}, 
traversable wormholes \cite{MT,MTY}, ``warp drive''\cite{WARP},
and possibly time machines \cite{MTY,AEVERETT}. As a result, much effort has 
been recently directed toward determining the extent to which
 quantum field theory permits violations of the local energy conditions. 
 One approach involves so-called 
``averaged energy conditions'' (see, for example, \cite{T}-\cite{VISSER}), 
i.e., averaging the local 
energy conditions over timelike or null geodesics. Another method employs 
``quantum inequalities'' (QI's) \cite{F78,FR95,F91}, which are  
constraints on the magnitude and duration of negative energy fluxes and 
densities. In a recent paper Ford and Roman \cite{FR96} have investigated the
 averaged weak(AWEC)
 and averaged null (ANEC) energy conditions, together with quantum inequalities
 on negative energy, in the context 
of evaporating black hole backgrounds in both two and four dimensions. They 
find that,  in all cases where the averaged energy 
conditions fail, there exist quantum inequalities which bound the magnitude and
 extent of the negative energy, and hence
 the degree of the violation.  In the four dimensional case, they first wrote 
down a QI-type bound by analogy with the 2D case, and  
then made use of the limited range of available numerical data outside the 
event horizon \cite{ELSTER}-\cite{JMO} to check whether the inequality 
is satisfied.
 However, generally, the energy integrals are to be evaluated over an infinite 
range to test the AWEC and ANEC\footnote{
and furthermore there seems to be an error in one set of the numerical data
\cite{JMO} they used, as pointed out by Visser\cite{VIS97}}. Therefore it 
remains interesting to see whether
 the qualitative 
features of the Ford-Roman results are still true if we can have analytical
 expressions for the energy-momentum tensor and employ them to evaluate the
 energy integrals over the whole range.  Fortunately, Visser \cite{VIS97} has
recently constructed a semi-analytical model for the stress-energy tensor
 corresponding to the conformally coupled massless scalar field that nicely
 fits all the known numerical data \cite{ELSTER}-\cite{JMO},  and satisfies all 
the known analytic features of the stress-energy tensor \cite{CF} in the Unruh 
vacuum describing a black hole evaporating into empty space.  It is worthwhile to point
out that the original Ford-Roman quantum inequality was proven for the minimally coupled,
 rather than the conformally coupled scalar field.
In this paper, we will use Visser's semi-analytical results to explicitly
 evaluate the relevant integrals to 
examine the energy conditions both outside and inside\footnote{ Although Visser's result
 is obtained using data from outside the event horizon there is nothing to prevent us from 
applying it inside.\cite{VIS97})} the black hole event horizon and test quantum 
 inequalities whenever the averaged energy conditions fail. It is worth noting that
 recently Pfenning and Ford have generalized the Ford-Roman quantum 
inequality-type restrictions on the stress-energy tensor for the minimally coupled 
scalar field in Minkowski spacetime to some static curved spacetimes.\cite{PFORD}

\section{ Averaged energy conditions and quantum inequalities}
Let us write  the four-dimensional Schwarzschild  metric in the form 

\begin{equation}
ds^2=-C dt^2+C^{-1}dr^2+ r^2d\Omega^2
\end{equation}
where $ C=1-2M/r$. Define a new variable $z=2M/r$. Then according to Visser's
 semi-analytical model,  the non-zero stress tensor components $T_{\mu\nu}$ 
(in Schwarzschild coordinates) are given by 

\begin{eqnarray}
T_{tt} &=&  p_\infty
  \Bigg[ \left({\xi\over20} - {k_5\over24} - {k_6\over20}\right) + 
          k_4 z^2 + 
          \left(-k_4 + {4 k_5\over3}\right) z^3 
\nonumber\\
&&\qquad 
       + \left({-\xi} -{5 k_5\over4} + {3 k_6\over2} \right)z^4 
       + \left( {9 \xi\over10} -{7 k_6\over5} \right)z^5 \Bigg],
\\
\nonumber\\
T_{rr} &=& -{z^2\over(1-z)^2} p_\infty
  \Bigg[ - \left({\xi\over20} - {k_5\over24} - {k_6\over20}\right) 
         + k_4  z^2 + 
          \left(-k_4 + {2 k_5\over3}\right) z^3 
\nonumber\\
&&\qquad 
       + \left(-{3 k_5\over4} + {k_6\over2} \right)z^4 
       + \left( { \xi\over10} -{3 k_6\over5} \right)z^5 \Bigg],
\\
\nonumber\\
T_{\theta\theta} &=& T_{\phi\phi}= p_\infty \; z^4 \; ( k_4 + k_5 z + k_6 z^2 ),
\\
\nonumber\\
T_{rt}  &=& T_{tr}= -\left( {\xi\over20}  - {k_5\over24} - {k_6\over20}\right)  \; p_\infty \; 
{z^2\over1-z},
\end{eqnarray}

where 
\begin{eqnarray}
 p_\infty =\frac{1}{90(16\pi)^2(2M)^4}\nonumber\\
k_4=26.5652,\quad k_5=-59.0214,\quad k_6=38.2068.\\
\end{eqnarray}
and $ T_{\mu\nu} $ is understood to denote the quantum expectation value in the Unruh vacuum 
state. In this paper, we only
consider the Unruh vacuum, as this is the vacuum which describes physically realistic
 collapse spacetimes in which the black hole forms at a finite time in the past.  
We shall evaluate AWEC and ANEC integrals along half-complete timelike and
null geodesics to determine where the averaged energy conditions fail and whether 
there exists any bound on the extent of violation. Usually the ANEC is defined using an integral over $(-\infty,
\infty)$, rather than over semi-infinite domains (i.e. over geodesics with past endpoints), as we shall consider here.
However, it is of some interest to study whether ANEC is satisfied in the spacetime of an object collapsing to form a 
black hole. The answer to this question might determine, for example, whether Penrose's singularity theorem \cite{P65} 
will still 
hold in the presence of local violations of the energy conditions, such as the Hawking evaporation process. In this case,
one would want ANEC to hold over half-complete null geodesics.

\paragraph*{1. Outgoing radial timelike observers}
The four-velocity of an outgoing radial timelike observer in 4D Schwarzschild spacetime is given by
\begin{equation}
u^{\mu}=(u^t,u^r)=\biggl({dt\over d\tau}, {dr\over d\tau},{d\theta \over d\tau},{ d\phi\over d\tau}
\biggr)=\biggl(\frac{k}{C},\sqrt{k^2-C},0,0\biggr)
\end{equation}
where $k$ is the energy per unit rest mass.
The energy density in this observer's frame is
\begin{equation}
T_{\mu\nu}u^{\mu}u^{\nu}=\frac{k^2}{C^2}(T_{tt}+C^2T_{rr})-CT_{rr}+\frac{2k\sqrt{k^2-C}}{C}\;T_{rt}
\end{equation}  

We consider, for simplicity, the case in which an observer is shot outward initially very close to the 
horizon at very large velocity, i.e., we are interested in the limits
\begin{equation}
\epsilon<<M, \qquad k>>1,
\end{equation}
where $r=2M+\epsilon $ is the observer's initial position.
Examine AWEC along the observer's trajectory to obtain
\begin{equation}
\int^\infty_{2m+\epsilon}\;T_{\mu\nu}u^{\mu}u^{\nu}\;d\tau=\int^\delta_0\;
\frac{2MT_{\mu\nu}u^{\mu}u^{\nu}}{z^2\sqrt{k^2-(1-z)}}
\;dz
\end{equation}
where $\delta= 1+\epsilon/2M. $ If we expand the integrand in inverse powers of $k$, then perform 
the integration using Maple  and expand the result in powers of $\epsilon$, we get
\begin{eqnarray}
\int^\infty_{2M+\epsilon}\;T_{\mu\nu}u^{\mu}u^{\nu}\;d\tau & \sim &
-\frac{k}{\eta M^2}\biggl( {2a_1\over15}\frac{1}{\epsilon}
-{a_2\over5}+O(\epsilon^0)\biggr) \nonumber\\
&&-\frac{a_3}{30\eta M^3}\frac{1}{k}+O(k^{-3})
\end{eqnarray}
where we have defined new constants 

\begin{eqnarray}
\eta=90\times(16)^3\pi^2, \quad a_1=6\xi-5k_5-6k_6=641.8662,
\nonumber\\
a_2 =6\xi-5k_5-6k_6=471.5354, \quad a_3=6\xi-20k_4-12k_6-15k_5=128.37324.
\end{eqnarray}

Plugging in the numerical values  and keeping only the leading order, we arrive at
\begin{equation}
\int^\infty_{2M+\epsilon}\;T_{\mu\nu}u^{\mu}u^{\nu}\;d\tau 
\sim -2.352\times10^{-5}{1\over M^2}{k\over\epsilon} \gprox 
-\frac{1}{M^2 \delta \tau}
\end{equation}
where
\begin{equation}
\delta\tau={\epsilon\over k}
\end{equation}
It is easy to show that near the event horizon,
\begin{equation}
T_{\mu\nu}u^{\mu}u^{\nu}\sim -{k^2\over{M^2(r-2M)^2}}
\end{equation}
 By the same argument as in Ref.\cite{FR96}, we see that $\delta\tau$ is the characteristic
 time scale for the changes in the energy density observed by the  outgoing radial timelike 
observers. Clearly,  AWEC is violated , but we still have a Ford-Roman type quantum inequality 
Eq.~(14) bounding the degree of 
violation seen by timelike geodesic observers who start out very close to the horizon at very high 
speed.  In this limit, the longer the timescale over which the observer sees a significant change in
 energy density, the smaller the magnitude of the integrated negative density seen by that observer.
Since the negative energy density drops off rapidly with increasing $r$,
to remain in the negative energy density the observer must stay close
to the horizon. The closer the observer is to the horizon, the larger
is the magnitude of the negative energy density.
However, in order to remain close to the horizon for a long time as seen
by a distant observer, the observer's trajectory must be nearly lightlike.
Therefore, although the observer spends a long time in the negative energy
region as seen by the distant observer, the {\it proper} time spent
in the region of appreciable negative energy (as measured by
$\delta \tau$) decreases with the observer's proximity to the horizon.

\paragraph*{2. Radial Null Geodesic}
ANEC is satisfied for ingoing radial null geodesics \cite{VIS97}, therefore we shall only examine
 the outgoing case.  Outgoing radial null geodesics are characterized by
\begin{equation}
K^{\mu}=E(C^{-1},1,0,0),
\end{equation}
where $E$ is an arbitrary positive constant, whose value fixes the scale of the affine parameter 
$\lambda. $
Thus, we have 
\begin{eqnarray}
T_{\mu\nu}K^{\mu}K^{\nu}&=&{E^2\over(1-z)}\biggl[C^{-1}T_{tt}+CT_{rr}+2T_{tr}\biggl]\nonumber\\
 &&= p_\infty {z^5\over(1-z)^2}
\Bigg[  
{2k_5\over3} - \left(\xi+{k_5\over2}-k_6\right) z +
\left({4\xi\over5} - {4k_6\over5}\right)z^2
\Bigg]
\nonumber\\
&&= p_\infty {z^5\over(1-z)^2}
\Bigg[
-\alpha-\beta z+\gamma z^2
\Bigg],
\nonumber\\
\end{eqnarray}
where 
\begin{equation}
\alpha={2k_5\over3}=39.3476, \quad \beta= \left(\xi+{k_5\over2}-k_6\right)=28.2825, 
\quad \gamma=\left({4\xi\over5} - {4k_6\over5}\right)=46.23456.
\end{equation}

Note that there is a discontinuity at the event horizon in the above expression, so we shall
 address the null rays on the horizon separately. It is easy to show that if 
\begin{equation}
r> \frac{4\gamma}{\beta +\sqrt{\beta^2+4\alpha\gamma}} M \sim 1.56523376 M
\end{equation}
then $T_{\mu\nu}K^{\mu}K^{\nu}$ is negative, therefore the local 
energy condition is also violated along the null geodesics slightly inside the horizon. 
Consider first  an outgoing null geodesic starting at $2M+\epsilon,$ with $ \epsilon<<M $, 
and examine the ANEC along this ray to get
\begin{equation}
I_1=\int^{\infty}_0\;T_{\mu\nu}K^{\mu}K^{\nu}d\lambda
 \sim -{2a_1\over15\eta M^2}{E\over\epsilon}+O(\epsilon^0).
\end{equation}
 let us now define 
\begin{equation}
\delta\lambda={\epsilon\over E}
\end{equation}
and note that $\delta\lambda$ 
is the characteristic affine parameter distance over which the negative energy is
decreasing along the outgoing geodesics.  Thus ANEC is violated along this ray, but we can rewrite Eq.~(21 )
 in the form of a QI
\begin{equation}
\int^{\infty}_0\;T_{\mu\nu}K^{\mu}K^{\nu}d\lambda 
\sim -2.3522\times 10^{-5}{1\over M^2{\delta\lambda}} 
\gprox -{1\over M^2{\delta\lambda}}
\end{equation}
This is the null version of Eq.~(14).
For outgoing rays inside the horizon starting at $r=2M-\epsilon$, we have

\begin{equation}
I_2=\int^{\lambda_f}_0\;T_{\mu\nu}K^{\mu}K^{\nu}d\lambda=
{1\over E}\int^{2M-\epsilon }_{r_{min}}\;T_{\mu\nu}K^{\mu}K^{\nu}dr\;,
\end{equation}
where $\lambda_f$ is the value of the affine parameter at $r=r_{min}$,
the minimum value of $r$ attained by the null rays.
If we perform the integration and expand the result in the small $r_{min}$ limit,  
with $\epsilon$ fixed, we get
\begin{equation}
I_2\sim {32(\xi-k_6)\over{5\eta}}\frac{ME}{r^4_{min}}+O(r^{-3}_{min})>0
\end{equation}
However, if we take the limit of $ \epsilon\rightarrow 0$ with $r_{min}$ fixed, then 
\begin{equation}
I_2\sim -{2a_1\over{15\eta M^2}}{E\over\epsilon}+O(\epsilon^0)<0
\end{equation}

Thus we see that ANEC is violated for outgoing null rays just outside the
horizon, but it is satisfied for outgoing null rays inside the horizon
in the limit $r_{min} \rightarrow 0$.
(The divergence in Eq.~(24) as $\epsilon \rightarrow 0$
may be circumvented by an appropriate choice of scaling of the affine
parameter, i.e., by an appropriate choice of $E$ \cite{FR96}.
 Outgoing rays which
originate in the region $1.56523376M<r<2M$ initially encounter negative energy,
and later encounter positive energy near the singularity, $r=0$. These 4D results would 
seem to suggest that the semiclassical effects of negative energy, in processes such as black
hole evaporation, might not invalidate the Penrose singularity theorem \cite{P65} before quantum 
gravity effects become significant. Our four-dimensional 
results here are essentially in accordance with the 2D results obtained by Ford and Roman except 
for the numerical constants and factors of $M^{-2}$. This may
 indicate that results in the 2D are good examples of what is going on in the 4D.

\paragraph*{3. Null geodesics on the horizon}

 The Schwarzschild coordinates for the outgoing null geodesics outside the horizon are singular
 on the horizon. To avoid this singularity, let us transform to the Kruskal null coordinates

\begin{equation}
U=-e^{-\kappa u}, \qquad V=e^{\kappa v}
\end{equation}
where $ u=t-r^*, v=t+r^*$,  with $r^*$ the usual tortoise coordinate and $\kappa=1/(4M)$  
the surface gravity. In thses coordinates the null vectors are
\begin{equation}
  K^{\bar\mu}=\biggl( {dU\over d\lambda},{dV\over d\lambda}\biggr)=
 \biggl( 0, {2E\over C}\kappa e^{\kappa v}\biggr)
\end{equation}
However, this is still not an affinely parameterized null tangent vector on the horizon, 
although the singularity can be removed by a proper choice of the scaling of $ E$ \cite {FR96}.  
The affinely parametrized null tangent vector on the horizon is \cite{WALDGR}
\begin{equation}
k^{\bar\mu}=(0,\kappa)
\end{equation}
 To examine the ANEC on the horizon, let us consider the integral
$\int_{}^{}\, T_{\bar \mu \bar \nu} \, k^{\bar \mu}\,k^{\bar \nu}\, dV$
along a portion of the future event horizon. Here $T_{\bar \mu \bar \nu}$
is the vacuum expectation value of the stress-tensor in the Unruh vacuum state
expressed in Kruskal null coordinates. The Kruskal
advanced time coordinate $V$, is
an affine parameter on the horizon ( \cite{WALDGR}, \cite{WALDQFT}). Using
Eq.~(29), and a straightforward coordinate transformation of
Eqs.~(2)-~(5), we obtain

\begin{equation}
\int_{V_0}^{\infty}\, T_{\bar \mu \bar \nu} \, k^{\bar \mu}\,k^{\bar \nu}\, dV
\,\sim \, - { a_4 \over {24\,\eta\,M^4\,V_0} } 
\sim -3.2493\times 10^{-7} {1\over M^4 V_0}\,.             \label{eq:Kint}
\end{equation}
where
\begin{equation}
a_4={6\xi\over 5}-k_5-{6k_6\over5}=28.37324.
\end{equation}

It is worth pointing out that for the Unruh vacuum describing a realistic collapse situation, 
 $V_0$ will not be zero. 
 There is a smallest allowed value of $V_0=V_{min}$, which is the value of the $V$=const. line at which the 
worldline of the surface of the collapsing star intersects the future event horizon.
So, the ANEC integral along a portion of the future event horizon is negative but finite.

To see if there is a Ford-Roman QI-type bound in this case, let us rescale the coordinates as follows
\begin{equation}
\hat U = {\kappa}^{-1}\,U\,, \hat V = {\kappa}^{-1}V\;.
\end{equation}
Then we find 
\begin{equation}
 T_{\hat \mu \hat \nu} \, k^{\hat \mu}\,k^{\hat\nu}\,= - { 2a_4 \over {3\,\eta\,M^2 \hat V^2} } \,
\end{equation}
Thus 

\begin{eqnarray}
\int_{\hat {V_0}}^{\infty}\, T_{\hat \mu \hat \nu} \, k^{\hat \mu}\,k^{\hat
\nu}
\, d{\hat V} \,& = & \, -
\frac{2a_4}{3\eta M^2} \int_{\hat {V_0}}^{\infty}\, \Biggl( {1 \over { {\hat V}^2} }
\Biggr)\, d{\hat V}                                        \nonumber \\
& = & \, - { 2a_4 \over {3 \eta M^2 \, \hat {V_0} } } 
\sim  - 2.5994\times 10^{-6} { 1 \over { M^2 \, \hat {V_0} } } \,.     
\end{eqnarray}

Obviously, $\hat V_0$ can be interpreted as the characteristic affine parameter distance over 
which the integrand falls off in the sense that if we increase $\hat V$
 from $\hat V = \hat {V_0}$ to $\hat V = 2 \hat {V_0}$,
$T_{\hat \mu \hat \nu} \, k^{\hat \mu}\,k^{\hat \nu}$ falls off to $1/4$ of
its initial value. Thus the interpretation of the QI for null geodesics
on the horizon is the same as that for Eq.~(23).

\section{Conclusions}

In this paper, we have examined, by explicitly evaluating the relevant integrals, AWEC and 
ANEC in four-dimensional evaporating Schwarzschild black hole spacetimes using Visser's 
semi-analytical model for the stress-energy tensor for the conformally coupled 
massless scalar field. We found that AWEC is violated for outgoing radial timelike geodesics 
which start just outside the horizon and reach infinity. Similarly, ANEC is violated for
 half-complete outgoing radial null geodesics which start just outside the horizon, and 
in these two cases, there exist QI-type bounds on the degree of violation. These confirm 
Ford-Roman's results. With the semi-analytical expressions for the stress-energy tensor 
components, we were able to further examine the ANEC on the horizon and ANEC for outgoing null 
rays inside the horizon. It was discovered that ANEC is satisfied along the outgoing null geodesics
 inside the horizon because of a large positive contribution to the integral near the singularity.
 This suggests that  the semiclassical effects of negative energy, in processes
 such as black  hole evaporation, might not invalidate the Penrose singularity \cite{P65} before 
quantum gravity effects become significant.  It is interesting to notice that 
 our 4D results are essentially in accordance with the 2D results obtained by 
Ford and Roman, and this would seem to suggest that 2D results may be a very 
good indicator of what is happening in the 4D in some cases.  However, it was 
found that ANEC is violated for 
radial null geodesics on the horizon and that there exists a similar Ford-Roman
 QI-type bound on
the degree of violation. It is worth pointing out that in our discussions it is assumed that
 the background metric is static, thus ignoring the backreaction of the Hawking radiation. 
This approach may well be valid during the early stages of black hole evaporation. However, 
since the Hawking evaporation is a highly dynamic process, sooner or later 
a point will be reached in which the backreaction can no longer be ignored. We then need 
to use an evolving background metric, such as the Vaidya metric, to model realistic evaporating
 black holes. Therefore it would be interesting to do a similar analysis using an evolving metric
 to examine the energy conditions and test quantum inequalities proposed by Ford
and Roman. Hopefully, we shall address this issue in the future.

\begin{acknowledgments}

I wish to thank Prof. Larry Ford and Prof. Tom Roman for their very helpful discussions, comments and 
their critical readings of the manuscript. I also wish to thank Prof. Tom Roman for suggesting 
this topic to me.

\end{acknowledgments}

\end{document}